\documentclass[aps,prd,onecolumn,groupedaddress,nofootinbib,amssymb]{revtex4}
\usepackage[T1]{fontenc}
\usepackage[latin1]{inputenc}
\usepackage{graphicx}
\usepackage[english]{babel}

\begin{document}

\title{\textbf{Stochastic background of relic scalar gravitational waves
from scalar-tensor gravity}}

\author{Salvatore Capozziello}
\affiliation{\emph{Dipartimento di Scienze fisiche,  Universit\`a
di Napoli {}`` Federico II'', INFN Sezione di Napoli, Compl. Univ.
di Monte S. Angelo, Edificio G, Via Cinthia, I-80126, Napoli,
Italy}}

\author{Christian Corda}
\affiliation{\emph{INFN Sezione di Pisa and  Universit\`a di Pisa,
Via F. Buonarroti 2,I-56127, Pisa, Italy European Gravitational
Observatory (EGO), Via E. Amaldi, I-56021 Pisa, Italy}}

\author{Mariafelicia De Laurentis}
\affiliation{\emph{Politecnico di Torino and INFN Sezione di Torino, Corso Duca
degli Abruzzi 24, I-10129 Torino, Italy}}

\begin{abstract}
A stochastic background of relic gravitational waves is achieved
by  the so called adiabatically-amplified zero-point fluctuations
process derived from  early inflation.  In principle, it provides
a distinctive spectrum of relic gravitational waves. In the
framework of scalar-tensor gravity, we discuss the  scalar modes
of gravitational waves and  the primordial production of this
scalar component which is generated beside  tensorial one. We
discuss  also the upper limit for such a  relic scalar component
with respect to the WMAP constraints.
\end{abstract}

\maketitle


\vspace{5.mm}

The design and the construction of a number of sensitive detectors
for gravitational waves (GWs) is underway today. There are some
laser interferometers like the VIRGO detector,  built in Cascina,
near Pisa, Italy, by a joint Italian-French collaboration, the GEO
600 detector  built in Hanover, Germany, by a joint Anglo-German
collaboration, the two LIGO detectors  built in the United States
(one in Hanford, Washington and the other in Livingston,
Louisiana) by a joint Caltech-Mit collaboration, and the TAMA 300
detector, in Tokyo, Japan. Many bar detectors are currently in
operation too, and several interferometers and bars are in a phase
of planning and proposal stages (for the current status of
gravitational waves experiments see \cite{Ligo1,Ligo2}). The
results of these detectors will have a fundamental impact on
astrophysics and gravitation physics. There will be a huge amount
of experimental data to be analyzed, and theorists would face new
aspects of  physics from such a data stream. Furthermore, GW
detectors  will  be of fundamental importance  to probe the
General Relativity or every alternative theory of gravitation
\cite{CT,Capozz,CC,Tobar}. A possible target of these experiments
is the so called stochastic background of gravitational waves
\cite{Allen,AO,Maggiore,Grishchuk,Corda}. The production of the
primordial part of this stochastic background (relic GWs) is well
known in literature starting from the works by \cite{Allen} and
\cite{Grishchuk}, which, using the so called
adiabatically-amplified zero-point fluctuations process, have
shown in two different ways how the  inflationary scenario for the
early universe can, in principle, provide the signature for the
spectrum of relic GWs.

In this letter, after a short summary of scalar-tensor gravity
admitting the existence of  GW scalar modes, the primordial
production of this scalar component is discussed. Besides, the
results are confronted to the WMAP data \cite{Bennet,Spergel} in
order to achieve an upper limit for the scalar part of relic GWs.

Scalar-tensor gravity theories are a particular case of Extended
Theories of Gravity  which are revealing a useful paradigm to deal
with several problems  in cosmology, astrophysics and fundamental
physics (for a comprehensive discussion see, for example
\cite{Odintsov,Mauro}).  In the most general case, considering
only the Ricci scalar among the curvature invariants,   they
arises from the action

\begin{equation}
S=\int d^{4}x\sqrt{-g}\left[F(R,\square R,\square^{2}
R,\square^{k}
R,\phi)-\frac{\epsilon}{2}g^{\mu\nu}\phi_{;\mu}\phi_{;\nu}+\mathcal{L}_{m}\right],\label{eq:
high-order}\end{equation} where $F$ is an unspecified function of
curvature invariants and of a scalar field $\phi$ and $\square$ is
the D'Alembert operator. The term $\mathcal{L}_{m}$ is the
minimally coupled ordinary matter contribution.  Scalar-tensor
gravity, is recovered from (\ref{eq: high-order}) through the
choice
\begin{equation}
\begin{array}{ccc}
F(R,\phi)=f(\phi)R-V(\phi), &\, &
\epsilon=-1\end{array}\,.\label{eq: ST}
\end{equation}
Considering  (\ref{eq: ST}), a general action for scalar-tensor
gravity in four dimensions is
\begin{equation}
S=\int
d^{4}x\sqrt{-g}\left[f(\phi)R+\frac{1}{2}g^{\mu\nu}\phi_{;\mu}\phi_{;\nu}-V(\phi)+\mathcal{L}_{m}\right]\,,
\label{eq: scalar-tensor}
\end{equation}
which can be recast  in a Brans-Dicke-like form \cite{BD} by
\begin{equation}
\begin{array}{ccc}
\varphi=f(\phi)\,, & \omega(\varphi)=\frac{f(\phi)}{2'f(\phi)}\,,
& W(\varphi)=V(\phi(\varphi))\\,,\end{array}\label{eq: scelta}
\end{equation}
and then
\begin{equation}
S=\int d^{4}x\sqrt{-g}\left[\varphi
R-\frac{\omega(\varphi)}{\varphi}g^{\mu\nu}\varphi_{;\mu}\varphi_{;\nu}-W(\varphi)+
\mathcal{L}_{m}\right]\,.\label{eq: scalar-tensor2}
\end{equation}

By varying the action (\ref{eq: scalar-tensor2}) with respect to
$g_{\mu\nu}$, we obtain  the field equations
\begin{equation}
\begin{array}{c}
G_{\mu\nu}=-\frac{4\pi\tilde{G}}{\varphi}T_{\mu\nu}^{(m)}+\frac{\omega(\varphi)}{\varphi^{2}}(\varphi_{;\mu}\varphi_{;\nu}-\frac{1}{2}g_{\mu\nu}g^{\alpha\beta}\varphi_{;\alpha}\varphi_{;\beta})+\\
\\+\frac{1}{\varphi}(\varphi_{;\mu\nu}-g_{\mu\nu}\square \varphi)+\frac{1}{2\varphi}g_{\mu\nu}W(\varphi)\end{array}\label{eq: einstein-general}\end{equation}
while the variation with respect to $\varphi$ gives the  Klein -
Gordon equation
\begin{equation}
\square
\varphi=\frac{1}{2\omega(\varphi)+3}\left[-4\pi\tilde{G}T^{(m)}+2W(\varphi)+\varphi
W'(\varphi)+\frac{d\omega(\varphi)}{d\varphi}g^{\mu\nu}\varphi_{;\mu}\varphi_{;\nu}\right].\label{eq:
KG}\end{equation} We are assuming physical units $G=1$, $c=1$ and
$\hbar=1$. $T_{\mu\nu}^{(m)}$ is the matter stress-energy tensor
and $\tilde{G}$ is a dimensional, strictly positive, gravitational
coupling  constant \cite{Capozz,CC}. The Newton constant is
replaced by the effective coupling
\begin{equation}
G_{eff}=-\frac{1}{2\varphi},\label{eq: newton eff}\end{equation}
which is, in general, different from $G$. General Relativity is
recovered for
\begin{equation}
\varphi=\varphi_0=-\frac{1}{2}.\label{eq: varphi}\end{equation} In
order to study gravitational waves, we  assume first-order, small
perturbations in vacuum ($T_{\mu\nu}^{(m)}=0$). This means
\begin{equation}
g_{\mu\nu}=\eta_{\mu\nu}+h_{\mu\nu}\,,\qquad
\varphi=\varphi_{0}+\delta\varphi\label{eq:
linearizza}\end{equation}
and
\begin{equation}
W\simeq\frac{1}{2}\alpha\delta\varphi^{2}\Rightarrow
W'\simeq\alpha\delta\varphi\label{eq: minimo}\end{equation} for
the self-interacting, scalar-field potential. These assumptions
allow to derive the "linearized" curvature invariants
$\widetilde{R}_{\mu\nu\rho\sigma}$ , $\widetilde{R}_{\mu\nu}$ and
$\widetilde{R}$  which correspond to $R_{\mu\nu\rho\sigma}$ ,
$R_{\mu\nu}$ and $R$, and then the linearized field equations
\cite{CC,Misner}

\begin{equation}
\begin{array}{c}
\widetilde{R}_{\mu\nu}-\frac{\widetilde{R}}{2}\eta_{\mu\nu}=-\partial_{\mu}\partial_{\nu}\Phi+\eta_{\mu\nu}\square \Phi\\
\\{}\square \Phi=m^{2}\Phi,\end{array}\label{eq: linearizzate1}\end{equation}
 where
\begin{equation}
\Phi\equiv-\frac{\delta\varphi}{\varphi_{0}}\,,\qquad
m^{2}\equiv\frac{\alpha\varphi_{0}}{2\omega+3}\,.\label{eq:
definizione}\end{equation} The case  $\omega=const$ and $W=0$ has
been analyzed in \cite{CC} considering the so-called
{}``canonical'' linearization  \cite{Misner}. In particular, the
transverse-traceless (TT) gauge (see \cite{Misner}) can be
generalized to scalar-tensor gravity obtaining the total
perturbation of a GW incoming in the $z+$ direction in this gauge
as
\begin{equation}
h_{\mu\nu}(t-z)=A^{+}(t-z)e_{\mu\nu}^{(+)}+A^{\times}(t-z)e_{\mu\nu}^{(\times)}+\Phi(t-z)e_{\mu\nu}^{(s)}.\label{eq:
perturbazione totale}\end{equation} The term
$A^{+}(t-z)e_{\mu\nu}^{(+)}+A^{\times}(t-z)e_{\mu\nu}^{(\times)}$
describes the two standard (i.e. tensorial) polarizations of a
gravitational wave arising from General Relativity in the TT gauge
\cite{Misner}, while the term $\Phi(t-z)e_{\mu\nu}^{(s)}$ is the
extension of the TT gauge to the scalar case. This means that, in
scalar-tensor  gravity, the scalar field generates a third
component for the tensor polarization of GWs. This is because
three different  degrees of freedom are present (see Eq.(32) of
\cite{CC}), while only two are present in standard General
Relativity.

Let us now take into account the primordial physical  process
which gave rise to a characteristic spectrum $\Omega_{sgw}$ for
the early stochastic background of relic scalar GWs. The
production physical process has been analyzed, for example, in
\cite{Allen,Grishchuk,Allen2} but only for the first two tensorial
components of eq. (\ref{eq: perturbazione totale}) due to standard
General Relativity. Actually the process can be improved
considering also the third scalar-tensor component.

Before starting with the analysis, it has to be emphasized that,
considering a stochastic background of scalar GWs, it can be
described in terms of the scalar field $\Phi$ and characterized by
a dimensionless spectrum (see the analogous definition for
tensorial waves in
\cite{Allen,AO,Maggiore,Grishchuk})\begin{equation}
\Omega_{sgw}(f)=\frac{1}{\rho_{c}}\frac{d\rho_{sgw}}{d\ln
f},\label{eq: spettro}\end{equation} where
 \begin{equation}
\rho_{c}\equiv\frac{3H_{0}^{2}}{8\pi G}\label{eq: densita'
critica}\end{equation} is the (actual) critical energy density of
the Universe, $H_0$ the today observed Hubble expansion rate, and
$d\rho_{sgw}$ is the energy density of the scalar part of the
gravitational radiation contained in the frequency range $f$ to
$f+df$. We are considering now standard units.

The existence of a relic stochastic background of scalar GWs is a
consequence of general assumptions. Essentially it derives from
basic principles of Quantum Field Theory and General Relativity.
The strong variations of gravitational field in the early Universe
amplifies the zero-point quantum fluctuations and produces relic
GWs. It is well known that the detection of relic GWs is the only
way to learn about the evolution of the very early universe, up to
the bounds of the Planck epoch and the initial singularity
\cite{Allen,AO,Maggiore,Grishchuk,Corda}. It is very important to
stress the unavoidable and fundamental character of such a
mechanism. It directly derives from the  inflationary scenario
\cite{Watson,Guth}, which well fit the WMAP data in particular
good agreement with almost exponential inflation and spectral
index $\approx1$, \cite{Bennet,Spergel}.

A remarkable fact about the inflationary scenario is that it
contains a natural mechanism which gives rise to perturbations for
any field. It is important for our aims that such a mechanism
provides also a distinctive spectrum for relic scalar GWs. These
perturbations in inflationary cosmology arise from the most basic
quantum mechanical effect: the uncertainty principle. In this way,
the spectrum of relic GWs that we could detect today is nothing
else but the adiabatically-amplified zero-point fluctuations
\cite{Allen,Grishchuk}. The calculation for a simple inflationary
model can be performed for the scalar field component  of eq.
(\ref{eq: perturbazione totale}). Let us assume that the early
Universe is described an inflationary de Sitter phase emerging in
a radiation dominated phase \cite{Allen,AO,Grishchuk}. The
conformal metric element is
\begin{equation}
ds^{2}=a^{2}(\eta)[-d\eta^{2}+d\overrightarrow{x}^{2}+h_{\mu\nu}(\eta,\overrightarrow{x})dx^{\mu}dx^{\nu}],\label{eq: metrica}\end{equation}
where, for a purely scalar GW the metric perturbation (\ref{eq: perturbazione totale})
reduces to
\begin{equation}
h_{\mu\nu}=\Phi e_{\mu\nu}^{(s)},\label{eq: perturbazione
scalare}\end{equation} Following \cite{Allen,Grishchuk}, in the de
Sitter phase, we have:
\\
\begin{center}
\begin{tabular}{|c|c|}
\hline $\eta<\eta_{1}$& conformal time\tabularnewline \hline
$P=-\rho$& equation of state\tabularnewline \hline
$\eta_{1}^{2}\eta_{0}^{-1}\left(2\eta-\eta\right)^{-1}$& scale
factor\tabularnewline \hline $H_{ds}=c\eta_{0}/\eta_{1}^{2}$&
Hubble constant \tabularnewline \hline
\end{tabular}
\end{center}
\[\]
while, in the radiation dominated phase we have, respectively,
\[\]
\begin{center}
\begin{tabular}{|c|c|}
\hline $\eta>\eta_{1}$& conformal time\tabularnewline \hline
$P=\rho/3$& equation of state\tabularnewline \hline
$\eta/\eta_{0}$& scale factor\tabularnewline \hline
$H=c\eta_{0}/\eta^{2}$& Hubble constant \tabularnewline \hline
\end{tabular}
\end{center}
\[\]
$\eta_1$ is the inflation-radiation transition conformal time and
$\eta_0$ is the value of conformal time today.  If we express the
scale factor in terms of comoving time $cdt=a(t)d\eta$,
 we have
\begin{equation}
a(t)\propto\exp(H_{ds}t)\label{eq: inflazione}\,,\qquad
a(t)\propto\sqrt{t}\label{eq: dominio radiazione}\end{equation}
 for the de Sitter and radiation phases respectively. In order to solve
 the horizon and flatness problems, the condition
${\displaystyle \frac{a(\eta_{0})}{a(\eta_{1})}>10^{27}}$ has to
be satisfied. The relic scalar-tensor GWs are the weak
perturbations $h_{\mu\nu}(\eta,\overrightarrow{x})$ of the metric
(\ref{eq: perturbazione scalare}) which can be written in the form

\begin{equation}
h_{\mu\nu}=e_{\mu\nu}^{(s)}(\hat{k})X(\eta)\exp(\overrightarrow{k}\cdot\overrightarrow{x}),\label{eq:
relic gravity-waves}\end{equation} in terms of the conformal time
$\eta$ where $\overrightarrow{k}$ is a constant wavevector. From
Eq.(\ref{eq: relic gravity-waves}), the scalar component is
\begin{equation}
\Phi(\eta,\overrightarrow{k},\overrightarrow{x})=X(\eta)\exp(\overrightarrow{k}\cdot\overrightarrow{x}).\label{eq:
phi}\end{equation} Assuming $Y(\eta)=a(\eta)X(\eta)$, from the
Klein-Gordon equation in the FRW metric, one gets
\begin{equation}
Y''+(|\overrightarrow{k}|^{2}-\frac{a''}{a})Y=0\label{eq:
Klein-Gordon}\end{equation} where the prime $'$ denotes derivative
with respect to the conformal time. The solutions of Eq. (\ref{eq:
Klein-Gordon})  can be expressed in terms of   Hankel functions
in both the
inflationary and radiation dominated eras, that is:\\
For $\eta<\eta_{1}$ \begin{equation}
X(\eta)=\frac{a(\eta_{1})}{a(\eta)}[1+H_{ds}\omega^{-1}]\exp-ik(\eta-\eta_{1}),\label{eq:
ampiezza inflaz.}\end{equation}
for $\eta>\eta_{1}$
\begin{equation}
X(\eta)=\frac{a(\eta_{1})}{a(\eta)}[\alpha\exp-ik(\eta-\eta_{1})+\beta\exp
ik(\eta-\eta_{1}),\label{eq: ampiezza rad.}\end{equation} where
$\omega=ck/a$ is the angular frequency of the wave (which is
function of the time being $k=|\overrightarrow{k}|$ constant),
$\alpha$ and $\beta$ are time-independent constants which we can
obtain demanding that both $X$ and $dX/d\eta$ are continuous at
the boundary $\eta=\eta_{1}$ between the inflationary and the
radiation dominated eras. By this constraint, we obtain
\begin{equation}
\alpha=1+i\frac{\sqrt{H_{ds}H_{0}}}{\omega}-\frac{H_{ds}H_{0}}{2\omega^{2}}\,,\qquad
\beta=\frac{H_{ds}H_{0}}{2\omega^{2}}\label{eq:
beta}\end{equation} In Eqs. (\ref{eq: beta}),
$\omega=ck/a(\eta_{0})$ is the angular frequency as observed
today, $H_{0}=c/\eta_{0}$ is the Hubble expansion rate as observed
today. Such calculations  are referred in  literature as the
Bogoliubov coefficient methods \cite{Allen,Grishchuk}.

\begin{figure}[ht]
\includegraphics[scale=0.9]{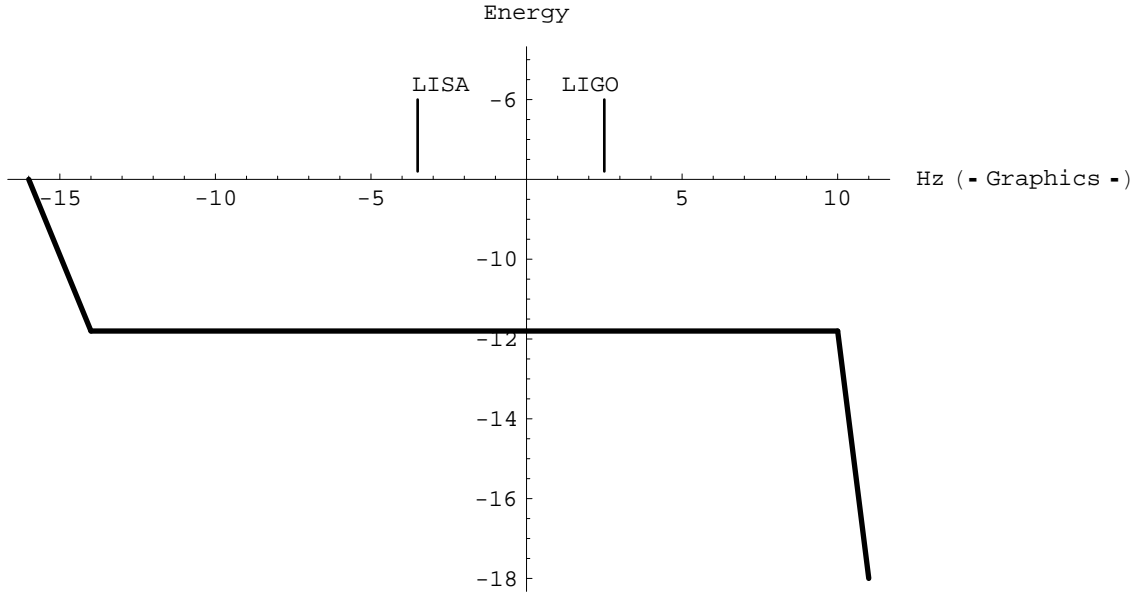}
\caption{The spectrum of relic scalar GWs in inflationary models
is flat over a wide range of frequencies. The horizontal axis is
$\log_{10}$ of frequency, in Hz. The vertical axis is
$\log_{10}\Omega_{gsw}$. The inflationary spectrum rises quickly
at low frequencies (wave which re-entered in the Hubble sphere
after the Universe became matter dominated) and falls off above
the (appropriately redshifted) frequency scale $f_{max}$
associated with the fastest characteristic time of the phase
transition at the end of inflation. The amplitude of the flat
region depends only on the energy density during the inflationary
stage; we have chosen the largest amplitude consistent with the
WMAP constrains on scalar perturbations. This means that at LIGO
and LISA frequencies, $\Omega_{sgw}<2.3*10^{-12}$}
\label{fig:spectrum}
\end{figure}

In an inflationary scenario, every  classical or macroscopic
perturbation is damped out by the  inflation, i.e. the minimum
allowed level of fluctuations is that required by the uncertainty
principle. The  solution (\ref{eq: ampiezza inflaz.}) corresponds
to a de Sitter vacuum state. If the period of inflation is long
enough, the today observable properties of the Universe  should be
indistinguishable from the properties of a Universe started in the
de Sitter vacuum state. In the radiation dominated phase, the
eigenmodes which describe particles are the coefficients of
$\alpha$, and these which describe antiparticles are the
coefficients of $\beta$ (see also \cite{tuning}). Thus, the number
of  particles created at angular frequency $\omega$ in the
radiation dominated phase is given by
\begin{equation}
N_{\omega}=|\beta_{\omega}|^{2}=\left(\frac{H_{ds}H_{0}}{2\omega^{2}}\right)^{2}.\label{eq:
numero quanti}\end{equation} Now it is possible to write an
expression for the energy density of the stochastic scalar relic
gravitons background in the frequency interval
$(\omega,\omega+d\omega)$ as
\begin{equation}
d\rho_{sgw}=2\hbar\omega\left(\frac{\omega^{2}d\omega}{2\pi^{2}c^{3}}\right)N_{\omega}=
\frac{\hbar
H_{ds}^{2}H_{0}^{2}}{4\pi^{2}c^{3}}\frac{d\omega}{\omega}=\frac{\hbar
H_{ds}^{2}H_{0}^{2}}{4\pi^{2}c^{3}} \frac{df}{f}\,,\label{eq: de
energia}\end{equation} where $f$, as above, is the frequency in
standard comoving time. Eq. (\ref{eq: de energia}) can be
rewritten in terms of the today and de Sitter value of energy
density being
\begin{equation} H_{0}=\frac{8\pi G\rho_{c}}{3c^{2}}\,,\qquad H_{ds}=\frac{8\pi G\rho_{ds}}{3c^{2}}.\end{equation}
Introducing the Planck density ${\displaystyle
\rho_{Planck}=\frac{c^{7}}{\hbar G^{2}}}$ the spectrum is given by
\begin{equation}
\Omega_{sgw}(f)=\frac{1}{\rho_{c}}\frac{d\rho_{sgw}}{d\ln
f}=\frac{f}{\rho_{c}}\frac{d\rho_{sgw}}{df}=\frac{16}{9}\frac{\rho_{ds}}{\rho_{Planck}}.\label{eq:
spettro gravitoni}\end{equation} At this point,  some  comments
are in order. First of all, such a calculation works for a
simplified model that does not include the matter dominated era.
If also such an era is also included, the redshift at equivalence
epoch has to be considered. Taking into account also results in
\cite{Allen2}, we get
\begin{equation}
\Omega_{sgw}(f)=\frac{16}{9}\frac{\rho_{ds}}{\rho_{Planck}}(1+z_{eq})^{-1},\label{eq:
spettro gravitoni redshiftato}\end{equation} for the waves which,
at the epoch in which the Universe becomes matter dominated, have
a frequency higher than $H_{eq}$, the Hubble parameter at
equivalence. This situation corresponds to frequencies
$f>(1+z_{eq})^{1/2}H_{0}$. The redshift correction in Eq.(\ref{eq:
spettro gravitoni redshiftato}) is needed since the today observed
Hubble parameter $H_{0}$ would result different  without a matter
dominated contribution. At lower frequencies, the spectrum is
given by \cite{Allen,Grishchuk}
\begin{equation}
\Omega_{sgw}(f)\propto f^{-2}.\label{eq: spettro basse
frequenze}\end{equation} As a further consideration, let us note
that the results (\ref{eq: spettro gravitoni}) and (\ref{eq:
spettro gravitoni redshiftato}), which are not frequency
dependent, does not work correctly in all the range of physical
frequencies. For waves with frequencies less than today observed
$H_{0}$, the notion of energy density has no sense, since the
wavelength becomes longer than the Hubble scale of the Universe.
In analogous way, at high frequencies, there is a maximal
frequency above which the spectrum rapidly drops to zero. In the
above calculation, the simple assumption that the phase transition
from the inflationary to the radiation dominated epoch is
instantaneous has been made. In the physical Universe, this
process occurs over some time scale $\Delta\tau$, being
\begin{equation}
f_{max}=\frac{a(t_{1})}{a(t_{0})}\frac{1}{\Delta\tau},\label{eq:
freq. max}\end{equation} which is the redshifted rate of the
transition. In any case, $\Omega_{sgw}$ drops rapidly. The two
cutoffs at low and high frequencies for the spectrum guarantee
that the total energy density of the relic scalar gravitons is
finite. For GUT energy-scale inflation it is of the order
\cite{Allen}
\begin{equation}
\frac{\rho_{ds}}{\rho_{Planck}}\approx10^{-12}.\label{eq: rapporto
densita' primordiali}\end{equation} These results can be
quantitatively constrained considering the recent WMAP release. In
fact, it is well known that WMAP observations put strongly severe
restrictions on the spectrum. In Fig. \ref{fig:spectrum} the
spectrum $\Omega_{sgw}$ is mapped : considering the ratio
$\rho_{ds}/\rho_{Planck}$, the relic scalar GW spectrum seems
consistent with the WMAP constraints on scalar perturbations.
Nevertheless, since the spectrum falls off $\propto f^{-2}$ at low
frequencies, this means that today, at LIGO-VIRGO and LISA
frequencies (indicated in fig. \ref{fig:spectrum}), one gets
\begin{equation} \Omega_{sgw}(f)h_{100}^{2}<2.3\times
10^{-12}.\label{eq: limite spettro WMAP}\end{equation} It is
interesting to calculate the  corresponding strain at $\approx
100Hz$, where interferometers like VIRGO and LIGO reach a maximum
in sensitivity. The well known equation for the characteristic
amplitude \cite{Allen,Grishchuk,Corda} adapted to the scalar
component of GWs can be used:
\begin{equation}
\Phi_{c}(f)\simeq1.26\times
10^{-18}(\frac{1Hz}{f})\sqrt{h_{100}^{2}\Omega_{sgw}(f)},\label{eq:
legame ampiezza-spettro}\end{equation} and then we obtain
\begin{equation}
\Phi_{c}(100Hz)<2\times 10^{-26}.\label{eq: limite per lo
strain}\end{equation}

Then, since we expect a sensitivity of the order of $10^{-22}$ for
the above interferometers at $\approx100Hz$, we need to gain four
order of magnitude. Let us analyze the situation also at smaller
frequencies. The sensitivity of the VIRGO interferometer is of the
order of $10^{-21}$ at $\approx10Hz$ and in that case it is
\begin{equation}
\Phi_{c}(100Hz)<2\times 10^{-25}.\label{eq: limite per lo
strain2}\end{equation} The sensitivity of the LISA interferometer
will be of the order of $10^{-22}$ at $\approx 10^{-3} Hz$ and in
that case it is
\begin{equation}
\Phi_{c}(100Hz)<2\times 10^{-21}.\label{eq: limite per lo
strain3}\end{equation} This means that a stochastic background of
relic scalar GWs could be, in principle, detected by the LISA
interferometer.

In summary, the above results point out that a further scalar
component of GWs, coming from Extended Theories of Gravity, should
be seriously considered in  the signal detection of
interferometers. As already discussed in \cite{tuning}, this fact
could constitute an independent  test for alternative theories of
gravity or a further probe for General Relativity capable of
ruling out other  theories.

\end{document}